\documentstyle[12pt]{article}
\begin{document}
\title{Renormalized QRPA and double beta decay: a critical analysis of 
double Fermi transitions}

 \author{Jorge G. Hirsch$^1$\thanks{e-mail: hirsch@fis.cinvestav.mx},
	 Peter O. Hess$^2$\thanks{e-mail: hess@roxanne.nuclecu,unam.mx} and
	 Osvaldo Civitarese$^3$\thanks{Fellow of the CONICET, Argentina;
	e-mail: civitare@venus.fisica.unlp.edu.ar}\\
{\small\it $^1$Departamento de F\'{\i}sica, Centro de Investigaci\'on
y de Estudios Avanzados del IPN,}\\
{\small\it A. P. 14-740 M\'exico 07000 D.F.}\\
 {\small\it$^2$Instituto de Ciencias Nucleares, Universidad Nacional
Aut\'onoma de M\'exico,}\\
{\small\it Apdo. Postal 70-543, M\'exico 04510 D.F.}\\
 {\small\it$^3$ Departamento de F\'{\i}sica, Universidad Nacional de La
Plata, }\\
{\small\it c.c. 67 1900, La Plata, Argentina.}
}
\maketitle

\begin{abstract}
The proton-neutron monopole Lipkin model, which exhibits
some properties which are relevant for
those double beta decay ($\beta \beta$) transitions mediated by the Fermi 
matrix elements, is solved exactly in the proton-neutron 
two-quasiparticle  space. The exact results are compared with the ones
obtained by using the Quasiparticle Random Phase
(QRPA) and renormalized QRPA (RQRPA) approaches.
It is shown that the RQRPA violates the Ikeda Sum Rule and that
this violation may be common to any extension of the QRPA where
scattering terms are neglected in the participant one-body operators as well
as in the Hamiltonian.
This finding underlines the need of additional developments before the 
RQRPA could be adopted as a reliable tool to compute $\beta \beta$ processes.
\end{abstract}

\noindent
PACS number(s): 21.60.Jz, 21.60.Fw, 23.40.Hc

\section{Introduction}

The nuclear double beta decay could provide evidence on the existence of
massive neutrinos and right handed weak currents \cite{Ver86}. This exciting
possibility has attracted many theoretical and experimental work in the
 last years \cite{Moe93}. In order to extract information about this new
physics the
data must be complemented with theoretical nuclear matrix
elements which are strongly suppressed.

Ten years ago it was realized
that the quasiparticle random phase approximation (QRPA), including a
particle-particle channel in the residual interaction, can reproduce the
experimentally determined two-neutrino double-beta decay
($\beta\beta_{2\nu}$) half-lives \cite{Vog86,Eng87,Civ87,Mut89}.
This step allowed the description of single
and double-beta decay transitions in many nuclei.
However it was soon recognized
that the ground state to ground state $\beta\beta_{2\nu}$ transition
amplitudes are extremely
sensitive to the force parameters, thus limiting the
predictive power of the theory. The breakdown of the QRPA approach,
for some critical values of the model parameters, made the
theoretical description of some cases particularly
difficult; i.e.  the $\beta\beta_{2\nu}$
decay of $^{100}Mo$ \cite{Vog86,Eng87,Civ91,Civ91b,Gri92}.

Recently the use of a correlated vacuum in the QRPA
 equation of motion, the so-called renormalized QRPA
(RQRPA) \cite{Har64,Row68}, has been
 reformulated \cite{Cat94} and applied to the $\beta\beta_{2\nu}$ decay
\cite{Toi95}. It was found
that the formalism is stable beyond the point where the QRPA collapses.
The RQRPA method requires the solution of coupled-non-linear
equations, instead of the usual eigenvalue problem of the QRPA.
However the physical consequences of this highly non-linear behaviour,
represented by the inclusion of some terms beyond the QRPA order of
approximation, have not been explored carefully.

In the present paper we will use a simple solvable model, which is an
extension of the Lipkin model \cite{Lip65},
to compare the exact,
the QRPA and the RQRPA approaches. It will be shown that
if one remains at the leading order of approximation, i.e. by retaining
two-quasiparticle terms in the
relevant Fermi or Gamow-Teller transition operators as well as
in the Hamiltonian, the RQRPA violates the Ikeda sum
rule.

\section{The model}

The model hamiltonian \cite{Kuz88,Mut92,Civ94a} consists in a single
 particle, a pairing term for protons and neutrons and a schematic
 charge-dependent residual interaction including particle-hole
 and particle-particle channels.
 It has been shown in a recent series of papers that this
 interaction, treated in the framework of the QRPA, is as good as a G-matrix
 constructed from the OBEP 
Bonn potential in reproducing single- and double-beta decay matrix elements.
 \cite{Civ94a,Civ94b,Civ95}.

The schematic hamiltonian reads
\begin{equation}
H = H_p + H_n + H_{res} \hspace{1cm}, \label{hamex}
\end{equation}
\noindent
where
\begin{equation}
\begin{array}{c}
H_p = \sum\limits_p e_p a^\dagger_p a_p - G_p S^\dagger_p S_p \hspace{1cm}
H_n = \sum\limits_n e_n a^\dagger_n a_n - G_n S^\dagger_n S_n\\
H_{res} = 2 \chi \beta_J^- \cdot \beta_J^+ 
          - 2 \kappa P_J^- \cdot P_J^+ \hspace{1cm}. 
\end{array}
\end{equation}

In the above expression the following definitions were introduced

\begin{equation}
\begin{array}{c}
S^\dagger_p = \sum\limits_p a^\dagger_p a^\dagger_{\bar p}/2 ,\hspace{1cm}
S^\dagger_n = \sum\limits_n a^\dagger_n a^\dagger_{\bar n}/2 ,
\\
\beta_J^- \cdot \beta_J^+ = \sum\limits_{M=-J}^J (-1)^M  :\beta^-_{JM}
(\beta^-_{J-M})^{\dagger}: \\
P_J^- \cdot P_J^+ = \sum\limits_{M=-J}^{J} (-1)^M  :P^-_{JM}
(P^-_{J-M})^{\dagger}:\\
 \beta^-_{JM} = \sum_{i,j} <i|{\cal O}_{JM} |j> a^{\dagger}_i a_j
 \hspace{1cm}
P^-_{JM} = \sum_{i,j} <i|{\cal O}_{JM} |j> a^{\dagger}_i
 a^{\dagger}_{\bar j}\\
{\cal O}_{1M} = \sigma_M \tau^- \hspace{1cm} {\cal O}_{00} = \tau^-
\end{array}
\end{equation}

\noindent
$ a^{\dagger}_p = a^\dagger_{j_p m_p}$ being the particle creation
 operator and
 $a^{\dagger}_{\bar p} = (-1)^{j_p -m_p} a^{\dagger}_{j_p -m_p}$ its time
reversal.
 
As mentioned above, the hamiltonian (\ref{hamex}) with $J=1$ provides a
reasonable description of the main physics involved in Gamow-Teller
transitions. The parameters $\chi$ and $\kappa$ play the role of the
renormalization factors $g_{ph}$ and $g_{pp}$ introduced in the
literature \cite{Vog86,Eng87,Civ87,Mut89}.

For the case of $J=0$ the hamiltonian (\ref{hamex}) can be reduced to an
isospin scalar if its parameters are selected as

\begin{equation}
e_p =e_n, \hspace{1cm}\chi=0,\hspace{1cm}G_p=G_n=4\kappa.
\end{equation}
 If $\chi \neq 0$ the isospin symmetry is broken in the particle-hole 
channel and for $4\kappa \neq G_p,G_n$ this symmetry is broken in the
particle-particle channel. A similar identification of
this isospin breaking 
mechanism can be performed for the case of realistic interactions with
renormalized proton-proton ($g^p_{pair}$), neutron-neutron ($g^n_{pair}$)
and proton-neutron particle-particle ($g_{pp}$) strengths 
\cite{Vog86,Eng87,Civ87,Mut89}.
For the Fermi transitions, like the ones contributing to the nuclear matrix
elements associated with the neutrinoless double-beta-decay mode, the value
$g_{pp} =1.0$ is usually adopted, while $g^p_{pair}$ and $g^n_{pair}$
vary from 0.9 to 1.2 to reproduce the observed proton and neutron pairing
gaps in medium and heavy mass nuclei. This parametrization, adopted for
realistic interactions, provides a first motivation for using the 
proton-neutron particle-particle strength $\kappa$ as a parameter 
independent of $G$.

In order to construct an exactly solvable model  to be used to test the
reliability of the RQRPA solutions for the single- and double-beta decay 
observables, we will make a strong approximation. We will consider a single
shell with the same angular momentum for protons and neutrons, i.e.
$j_p = j_n = j$. Adopting this model space is equivalent to work with
the one-level
limit of (\ref{hamex}). It will be shown that this model, which is not
intended to reproduce actual nuclear properties, does have the qualitative
features of a realistic pn-QRPA calculation. Indeed, excitation energies, 
single- and double-beta decay transition amplitudes and ground state
correlations depend on the particle-particle strength parameter $\kappa$
in the same way as they do in more elaborated calculations with many single
particle levels and with more realistic interactions. The advantage of such
a simplification lies upon the fact that the proton-neutron excitations
can be described within the framework of an  exactly solvable model. 
Particularly, the correspondence between the simplified version (one-shell
limit) of (\ref{hamex}) and the ordinary monopole Lipkin model\cite{Lip65}
 can be established if one chooses the channel with $J=0$ of hamiltonian 
(\ref{hamex}).
Physically, this case will correspond to Fermi type transitions but we
should emphasize the fact that the study of the model and not the
adjustment
of a given decay channel constitutes the essential aspect of the present
discussion.

In the single shell case, and after performing the Bogolyubov 
transformations, separately, for protons and neutrons, the first
two terms in (\ref{hamex}) become diagonal, i.e.
\begin{equation}
H_p = \epsilon_p \sum\limits_{m_p} \alpha^\dagger_{pm_p} \alpha_{pm_p},
\hspace{1cm}
H_n = \epsilon_n \sum\limits_{m_n} \alpha^\dagger_{nm_n} \alpha_{nm_n},
\end{equation}

\noindent
being $\epsilon_p, \epsilon_n$ the quasiparticle energies and 
$\alpha^\dagger_p, \alpha^\dagger_n$ the quasiparticle creation
operators \cite{Row70}.

The linearized hamiltonian, 
neglecting the scattering terms ($\alpha^\dagger_p \alpha_n,
\alpha^\dagger_n \alpha_p$) which do not
contribute at the QRPA order, reads

\begin{equation}
\begin{array}{ll}
H = \epsilon_p &\sum\limits_{m_p}  \alpha^{\dagger}_{pm_p} \alpha_{pm_p}
 ~+~
  \epsilon_n \sum\limits_{m_n}  \alpha^{\dagger}_{nm_n} \alpha_{nm_n} ~+ 
\nonumber\\ &\\
 &2 \chi (2j+1)
\left[(u_p^2 v_n^2 + v_p^2 u_n^2 ) A^\dagger \cdot  A  ~+ \right.\nonumber\\
&\hspace{1cm}\left.u_p v_n v_{p} u_{n} A^\dagger \cdot A^\dagger  +
v_p u_n u_{p} v_{n}  A \cdot  A \right ] ~- \\ &\\
 &2 \kappa   (2j+1) \left
[(u_p^2 u_n^2 + v_p^2 v_n^2 ) A^\dagger \cdot A  ~- \right.\nonumber\\
&\hspace{1cm}\left. u_p u_n v_{p} v_{n} A^\dagger  \cdot A^\dagger
  - v_p v_n u_{p} u_{n}  A \cdot   A  \right ].
\end{array}  \label{hamori}
\end{equation}
\noindent
with
$$
A^\dagger   = \left [ \alpha^{\dagger}_p \otimes
\alpha^{\dagger}_n \right ]^{J=0}_{M=0} 
$$

To show that it is possible to reduce (\ref{hamori}) to an holomorphic
version of the Lipkin model we have considered, for simplicity,
one single particle orbital for protons and
neutrons with the same quasiparticle energies 
$\epsilon_p = \epsilon_n = (2j+1) G_i/4,~i=p,n$\cite{Row70}.

Under these approximations the hamiltonian (\ref{hamori}) has the form

\begin{equation}
H = \epsilon ~C + \lambda_1 A^\dagger  A + \lambda_2
( A^\dagger A^\dagger + A A ) \label{lipham}
\end{equation}
with

\begin{equation}
\begin{array}{l}
C \equiv \sum\limits_{m_p}  \alpha^{\dagger}_{pm_p} \alpha_{pm_p} +
    \sum\limits_{m_n} \alpha^{\dagger}_{nm_n} \alpha_{nm_n} ~,
\hspace{1cm} \hbox{and}
\nonumber \\
~\\
\lambda_1 = 2 \left [\chi (u_p^2 v_n^2 + v_p^2 u_n^2) -
\kappa (u_p^2 u_n^2 + v_p^2 v_n^2 ) \right ] ~,\hspace{.5cm}
 \lambda_2 = 2 ( \chi + \kappa ) u_p v_p u_n v_n ~.\nonumber
\end{array}
\end{equation}

 The operators $\{A, A^\dagger, C\}$ satisfy the SU(2) quasispin algebra
\cite{Rin80}

\begin{equation}
\left [ A,  A^\dagger \right ] = 1 - C /(2j+1),
\hspace{1cm}
\left [ C,  A^\dagger \right ] = 2  A^\dagger
\end{equation}

 A form similar to (\ref{lipham}) was introduced by Lipkin {\em et
 al.} in their original paper \cite{Lip65}. The usual Lipkin model is
obtained setting $\lambda_1 = 0$, since this term
 essentially renormalizes the single particle energy $\epsilon$.
 In our
 case it corresponds to the particular case $Z = N = (2j+1)/2$,
 which implies $v_p = v_n = u_p = u_n$, and $\chi = \kappa$.
However, we shall keep this term in order to be as close as possible
to the realistic situation.

\section{Exact solutions}

 To obtain the exact solutions of (\ref{lipham}) we have performed a 
Holstein-Primakoff mapping
 \cite{Rin80,Kle91} of this hamiltonian.
 It involves the substitution of
pair of fermions by functions of the exact boson
operators $b^\dagger $ and $ b$, which fulfill the exact
commutation rule $[b, b^\dagger ] = 1$. The relations between both set
of operators are the following

\begin{equation}
 A^\dagger \rightarrow b^\dagger  \left ( 1 - {\frac {b^\dagger b}
{(2j+1)} }\right )^{1/2}~,\hspace{.5cm}
 A \rightarrow \left ( 1 - {\frac {b^\dagger b} {(2j+1)} }\right )^{1/2}
b ~,\hspace{.5cm} C \rightarrow 2 b^\dagger b ~.
\end{equation}

 The exact solutions are obtained using the boson basis
\begin{equation} 
|n_b) = {\frac {(b^\dagger )^{n_b} } {\sqrt{n_b!}} } |),
\hspace{1cm}
b|) = 0.
\end{equation}

 The hamiltonian  (\ref{lipham}) has terms which change the number of 
 bosons in two units, thus the exact wave function does not have a definite
 number of bosons. At the same time, states with odd and even number of
 bosons are not connected. To construct a decay scheme for double Fermi
transitions the exact ground
state of the even-even nuclei will be represented by
the lowest energy state with an even
number of bosons while the $0^+$ states of the odd-odd nuclei are those
with an odd number of bosons.
Spurious states are avoided by
limiting the number of bosons to $0 \leq n_b \leq 2j+1$.
The physical states  are written

\begin{equation}
|\lambda_{even-even}) = \sum\limits_{n_b~even}^{2j+1} C^{\lambda_e}_{n_b}
 | n_b)~ ,\hspace{1cm}
|\lambda_{odd-odd}) = \sum\limits_{n_b~odd}^{2j+1} C^{\lambda_o}_{n_b} | n_b)~.
\end{equation}

It must be clear that what we call ``exact solutions'' are the exact 
solutions of the hamiltonian (\ref{lipham}), where the terms of the
form $\alpha^\dagger_p \alpha_n, \alpha^\dagger_n \alpha_p$ (scattering
terms) have been neglected. These exact solutions exhaust any extension
of the QRPA intended to diagonalize hamiltonian (\ref{lipham}). 
However, the solutions of the complete eigenvalue problem of the hamiltonian
(\ref{hamex}) span a larger Hilbert space, including states orthogonal 
to those present in the actual ``exact'' solutions. The operator 
$\beta^\pm$ is sensitive to this truncation of the Hilbert space generated by
solving the restricted problem defined by the use of hamiltonian 
(\ref{lipham}). While at the QRPA level it has no effect, in any extensions
beyond QRPA the transition amplitudes exhibit the missed components. 
Its effects on the Ikeda sum rule are described below.

\section{QRPA and RQRPA}

 The QRPA hamiltonian $H_{QRPA}$ can be obtained from Eq. (\ref{lipham})
by taking the limit $(2j+1) \rightarrow \infty$. It is given by
 \begin{equation}
H_{QRPA} = (2 \epsilon + \lambda_1) ~b^\dagger b ~+
\lambda_2 \{ ~b^\dagger b^\dagger ~+ ~b b \}.
\end{equation}

The QRPA states are generated with the one-phonon operator
 $O^\dagger_{QRPA} = X A^\dagger - Y A$ acting over the correlated QRPA
 vacuum $|0>$.
The quasiboson approximation assumes the $<0| [A,A^\dagger ]|0> = 1$,
leading to the normalization condition $X^2 - Y^2 = 1$.
The QRPA matrix is just a $ 2 \times 2$ one, with sub-matrices
$ {\cal A_{QRPA}} =  2 \epsilon + \lambda_1$ and ${\cal B_{QRPA}} = 2
\lambda_2$.
The eigen-energy is $ E_{QRPA} = [(2 \epsilon + \lambda_1)^2 - 4
\lambda_2^2]^{1/2}$. It becomes a pure imaginary number if $2 \lambda_2
 > 2 \epsilon + \lambda_1$.
It means that for this limit the zero-boson component of ground state
ceased to be dominant.

In the renormalized QRPA the structure of the ground state is
included explicitly \cite{Row68}, in the form 
\begin{equation}
|0> = {\cal N} e^S |BCS>~, \hspace{.5cm}S = {\frac {c A^\dagger A^\dagger}
{2 <0| [A,A^\dagger ]|0> } }\hspace{.3cm}.
\end{equation}

The RQRPA one-phonon state is given by
\begin{equation}
 O^\dagger_{RQRPA} |0> = \left [{\cal X} A^\dagger - {\cal Y} A \right ]
/ <0| [A,A^\dagger ]|0>^{1/2} |0> \end{equation}

 The condition $ O_{RQRPA} |0> = 0$ leads to the value
$c = {\cal Y/X}$. After some algebra
it is possible to show that $<0| [A,A^\dagger ]|0>  \equiv D
 = 1 - {\frac {2 {\cal Y}^2 D} {2j+1} } $ \cite{Cat94,Toi95}, and that
\begin{equation}
 D = \left [ 1 + {\frac {2 {\cal Y}^2 } {2j+1} } \right ]^{-1}.
\end{equation}

 The RQRPA submatrices are ${\cal A_{RQRPA}} =	2 \epsilon + \lambda_1 D$
 and ${\cal B_{RQRPA}} = 2 \lambda_2 D$. Given	$0 \leq D \leq 1$, the
 presence of $D$ multiplying both $\lambda_1$ and $\lambda_2$  gives the
needed reduction of the residual interaction  to avoid the collapse of
the QRPA equations \cite{Toi95}.
Due to this fact the RQRPA energy $E_{RQRPA}$ is always real. Its
value must be obtained by solving simultaneously the non-linear equations
for $E_{RQRPA}, {\cal X,Y}$ and $D$, which in the general case will
include all possible values of $J$ \cite{Toi95}.

\section{Results and discussion}

In the following we will present numerical results which correspond
to the model space and parameters

\begin{equation}
j = 9/2~,~~Z = 4~,~~ N = 6~,~~ \epsilon = 1 MeV~.
\end{equation}
In order to avoid dealing with small numbers, we now redifine the two 
parameters
\begin{equation}
\kappa \rightarrow \kappa' \equiv (2j+1) \kappa, \hspace{1cm}
\chi \rightarrow \chi' \equiv (2j+1) \chi.
\end{equation}
We selected the values $\chi' = 0$ or $0.5$.
The particle-particle strength $\kappa'$ is kept as a variable.

\subsection{Excitation energies}

 In Fig. 1 the excitation energies $E_1 -E_0, ~E_{QRPA}$ and $E_{RQRPA}$
 are plotted against $\kappa'$ for $\chi' =0$ (upper figure) and $\chi' =
 0.5$ (lower figure). In both cases the collapse of the QRPA is evident
at $\kappa' \simeq 1$. The RQRPA excitation  energy remains real but it
 decreases as compared with the exact one. For a relatively
large value of $\kappa'$,
 ($\kappa' \simeq 2$), the ground state and the first excited one tend to
become degenerate.

{\bf Fig. 1}

This figure strongly resemble Fig. 1 of ref. \cite{Toi95} and Fig. 2 of ref. 
\cite{Civ91b}, where the energy of the first excited state is plotted 
against the particle-particle strength parameter $g_{pp}$. The curves for the
QRPA and the RQRPA are quite similar to those shown here. The advantage
of the present simple model is that we can compare them with the exact
results, while in the general case with multiple single particle levels
the exact solutions are unknown.

\subsection{Boson number}

 The expectation value of boson number operator in the ground state
 measures the difference between the $QRPA$ ( or $RQRPA$) ground states
and the BCS vacuum. Its expected value should be half the number of
quasiparticle and for the different cases it is written as
\begin{equation}
\begin{array}{l}
<0|\hat n|0>_{exact}  = \sum\limits_{n~even}^{2j+1} |C^0_n|^2 n   ~, \\
~\\
 <0|\hat n|0>_{RPA} = Y^2 ~, \hspace{1cm}<0|\hat n|0>_{RQRPA} = {\cal Y}^2~. 
\end{array}
\end{equation}

Fig. 2 shows the behavior of this average number for $\chi' = 0$ and $0.5$.
Up to $\kappa' \simeq 0.7 $ the three quantities are
nearly indistiguishable.
 From there on, the QRPA overestimates this correlations, as pointed out
long ago \cite{Rin80}, and quickly collapses. The RQRPA does not
collapse but it shows about twice the exact number of bosons.

{\bf Fig. 2}

This behaviour will affect the Ikeda Sum Rule, as it will be shown below.

\subsection{$\beta$ and $\beta\beta$ transition amplitudes}

The Fermi $\beta^\pm$ operators in the quasiparticle basis are
\begin{equation}
\begin{array}{c}
\beta^- = \sqrt{2j+1} \left[ u_p v_n A^\dagger + v_p u_n A
- u_p u_n B^\dagger + v_p v_n B \right],\\
\beta^+ = (\beta^-)^\dagger, \hspace{1cm}
B^\dagger = [\alpha^\dagger_p \alpha_{\bar n}]^{J=0}.
\end{array}
\end{equation}

Neglecting the scattering terms $B^\dagger, B$, as it was done to obtain
hamiltonian (\ref{lipham}), leads to the QRPA order of approximation.
In this case the amplitude of the Fermi transitions connecting the 
ground state $|0)$ of the initial even-even nuclei and the ground 
and excited states $|0_\lambda>$ in the odd-odd nuclei are
\begin{equation}
\begin{array}{l}
 <0_\lambda|\beta^-|0>_{exact} ~= ~\sum\limits_{n~even}^{2j+1} \sqrt{2j+1}
 ~C^0_n \left ( u_p v_n C^{*\lambda}_{n+1} \left [ (1-{\frac n
{2j+1}})~(n+1) \right ]^{1/2} + \right.\\ \hspace{7cm} \left.
 v_p u_n C^{*\lambda}_{n-1} \left [ (1-{\frac n {2j+1}})~n \right ]^{1/2}
\right ) \\
<0_1|\beta^-|0>_{RPA} ~= ~\sqrt{2j+1} ~\left ( X ~u_p v_n + Y ~v_p u_n
\right ) \\
 <0_1|\beta^-|0>_{RQRPA} ~= ~\sqrt{2j+1} ~\left ( {\cal X}~ u_p v_n + {\cal
Y} ~v_p u_n \right ) \end{array}
\end{equation}

 Similar expressions hold  for $\beta^+$ interchanging the $u$'s and
 $v$'s. There are several states in the odd-odd nuclei which can be
 connected to the ground state of the even-even one (five in our
example). Only the state with the lowest energy, labeled with
 $|0_1>$, is described by the QRPA and the RQRPA.

 In order to study the $\beta\beta_{2\nu}$ decay amplitudes $M_{2\nu}$ in
this simple model we made the approximation $<0_{final}| \beta^- |0_\lambda>
 \approx <0_{initial}| \beta^- |0_\lambda> = <0_\lambda |\beta^+ |0>$
\cite{Vog86,Eng87}. In this way, we use the expression

\begin{equation}
 M_{2\nu} ~=~ \sum_\lambda {\frac {<0_\lambda |\beta^+ |0> <0_\lambda
|\beta^- |0> } {E_\lambda + \Delta} }
\end{equation}

 We selected $\Delta = 0.5~MeV$. The sum runs over all the odd-odd
states, which are only one in the QRPA and RQRPA.

Fig. 3 shows $M_{2\nu}$ as a function of $\kappa'$ for $\chi' = 0$ and $0.5$.
 Their behaviour is very similar of that encountered in realistic
 calculations \cite{Vog86,Civ87,Mut89,Toi95}, including its cancellation
 near the collapse in the QRPA description. The success of the RQRPA in
extending this curve far beyond the value of $\kappa'$ where the
 QRPA collapsed is clearly seen. By the other side,
 compared with the exact values of $M_{2\nu}$, the RQRPA performance
 is poor.
The overestimation of the ground state correlation
causes a premature cancellation of the $\beta\beta_{2\nu}$ transition
amplitude, as compared with the exact result.

{\bf Fig. 3}

\subsection{Ikeda Sum Rule}

There is an additional point which must be mentioned. The Ikeda Sum Rule
\begin{equation}
S^-  ~- ~S^+ ~= ~N -Z ~, \hspace{1cm}
S^\pm = \sum\limits_\lambda |<0_\lambda |\beta^\pm |0>|^2
\end{equation}
\noindent
must be fulfilled in any model where the Hilbert space for
the odd-odd nuclei includes all the states which can be connected to the
ground state via the beta decay.

{\bf Fig. 4}

The behaviour of $S^- - S^+$ against $\kappa'$ is exhibited in Fig. 4.
 As it is well known, in the QRPA this sum rule is always fulfilled,
 irrespective of how large the ground state correlations are. It is seen
in the figure as a dotted straight line, interrupted where the QRPA
collapses.
 The RQRPA strongly violates the sum rule by nearly 50 \% at large values
of $\kappa$.
The exact solution of (\ref{lipham}) also
violates the sum rule but in a smaller amount.
The origin of this problem can be attributed to the adopted structure
of both the hamiltonian and the transition operators. As it was mentioned
above neglecting the scattering terms in the 
operators H, $\beta^-$ and $\beta^+$ 
provides a plausible explanation of this failure. At the QRPA level these 
terms play no role,
but when ground states correlations are explicitly taken into account
they become relevant and cannot be neglected, neither in the transition
operators nor in the Hamiltonian.

\section{Conclusions}

 We have presented a solvable model which is holomorphic
to the Lipkin model and used it
to study the  QRPA and RQRPA approaches. The case of
double beta decay transitions of the Fermi type has been discussed.
 We have found that the excitation energies
and double beta decay amplitudes, for these transitions,
have the same qualitative behaviour found in realistic calculations.
It is shown that, as expected, the RQRPA does not collapse as the
QRPA description does when renormalized particle-particle correlations
are included. However, the apparent stability of the RQRPA is hindered
by the fact that it
strongly overestimates the effect of ground state correlations.
This tendency is reflected by the premature change in the
sign of $M_{2\nu}$, as compared to the exact solution, and
by the violation of the Ikeda Sum Rule.
In our opinion it implies that if one wants to
include  explicitly the so-called ground-state-correlations
in the fashion of \cite{Row68} one should also exceed
the two-quasiparticle order. The usually neglected
one-quasiparticle terms are included
in the RQRPA when the exact commutation relations for $A^\dagger$ and
 $A$ are used but they should be also included in the transition
operators and in the hamiltonian.
In our exact model, the inclusion of these terms implies the use
of the SO(5) algebra \cite{Kle91,Hec65} instead of the
present SU(2) one. Details about the work in progress will be
published elsewhere. \cite{Hir96}

\section{Acknowledgments}

Useful discussions with O. Casta\~nos, J. Engel and  S. Pittel are
acknowledged. One of the authors (J.G.H) thanks A. Mariano for useful
comments. This work was partially done during a visit
of J.G.H. and P.O.H. to the Institute of Nuclear Theory at the
University of Washington in Seattle. Partial
support of the Conacyt, the CONICET and the J.S. Guggenheim
Foundation is also acknowledged.

\newpage

\centerline{\bf Figure Captions}

\bigskip
Fig. 1: Excitation energy {\em vs.} the particle-particle strength $\kappa'$.
The particle-hole strength is $\chi' = 0$ (upper figure) and  $\chi' = 0.5$. The
exact solution is plotted with thin line, the QRPA with dots and the RQRPA with
a thick line.

\bigskip
Fig. 2: Number of bosons $<0|\hat n|0>$ in the ground state {\em vs.}
$\kappa'$.  Conventions are the same as in Fig. 1.

\bigskip
Fig. 3: $\beta\beta_{2\nu}$ transition amplitudes $M_{2\nu}$ {\em vs.}
$\kappa'$.  Conventions are the same as in Fig. 1.

\bigskip
Fig. 4: Ikeda sum rule {\em vs.} $\kappa'$. Conventions are
the same as in Fig. 1., for $\chi' = 0.5$ only.

\newpage

\begin{figure}
\setlength{\unitlength}{0.240900pt}
\ifx\plotpoint\undefined\newsavebox{\plotpoint}\fi
\sbox{\plotpoint}{\rule[-0.200pt]{0.400pt}{0.400pt}}%


\end{figure}
\vspace{1cm}
\centerline{Figure 4}

\end{document}